\documentclass[balance,upint,subscriptcorrection,varvw,mathalfa=cal=boondoxo,spanish,french,vietnamese,russian,greek,pdf-a,fontspec,colorlinks]{asmeconf}

\usepackage{float}
\usepackage{tikz}
\usepackage{siunitx}
\usepackage[none]{hyphenat}
\hypersetup{%
	pdfauthor={Adil Rasheed},
	pdftitle={Digital Twin for Wind Energy: Latest updates from the NorthWind project},                  
	pdfkeywords={Digital Twin, Wind Energy},
}

\begin{document}
	

	\title{Digital Twin for Wind Energy: Latest updates from the NorthWind project} 
	
	\SetAuthors{%
		Adil Rasheed\affil{1}\affil{2}\CorrespondingAuthor{},
        Florian Stadtmann\affil{1},
        Eivind Fonn\affil{2},
        Mandar Tabib\affil{2},
        Vasileios Tsiolakis\affil{3},
        Balram Panjwani\affil{4}, 
        Kjetil Andre Johannessen\affil{2}, 
		Trond Kvamsdal\affil{2}\affil{3},
        Omer San\affil{5},
        John Olav Tande\affil{6},
        Idar Barstad\affil{7},
        Tore Christiansen\affil{8},
        Elling Rishoff\affil{8},
        Lars Frøyd\affil{9},
        Tore Rasmussen\affil{10}
	}
	
\SetAffiliation{1}{Department of Engineering Cybernetics, NTNU, O.S. Bragstadsplass 2, 7034, Trondheim, Norway}
\SetAffiliation{2}{Mathematics and Cybernetics, SINTEF Digital, Klæbuveien 153, 7031, Trondheim, Norway}
\SetAffiliation{3}{Department of Mathematical Sciences, NTNU, Alfred Getz vei 1, 7034, Trondheim, Norway}
\SetAffiliation{4}{Metal Production and Processing, SINTEF Industry, S. P. Andersens veg 15B, 7031 Trondheim, Norway}
\SetAffiliation{5}{University of Tennessee Knoxville, Knoxville, TN 37996, USA}
\SetAffiliation{6}{SINTEF Energy Research, P.O. Box 4761 Torgarden, 7465 Trondheim, Norway}
\SetAffiliation{7}{Norconsult, Vestfjordgaten 4, 1338 Sandvika, Norway}
\SetAffiliation{8}{DNV, Veritasveien 1, 1363 Høvik, Norway}
\SetAffiliation{9}{4subsea, Hagaløkkveien 26, 1383 Asker, Norway}
\SetAffiliation{10}{ANEO, Klæbuveien 118, 7031, Trondheim, Norway}
	
	
	\maketitle
	\begingroup
	\renewcommand
	\thefootnote{\textsection}
	\endgroup
	
	
	
	\normalfont
	

\begin{abstract}
NorthWind, a collaborative research initiative supported by the Research Council of Norway, industry stakeholders, and research partners, aims to advance cutting-edge research and innovation in wind energy. The core mission is to reduce wind power costs and foster sustainable growth, with a key focus on the development of digital twins. A digital twin is a virtual representation of physical assets or processes that uses data and simulators to enable real-time forecasting, optimization, monitoring, control and informed decision-making. Recently, a hierarchical scale ranging from 0 to 5 (0 - Standalone, 1 - Descriptive, 2 - Diagnostic, 3 - Predictive, 4 - Prescriptive, 5 - Autonomous has been introduced within the NorthWind project to assess the capabilities of digital twins. This paper elaborates on our progress in constructing digital twins for wind farms and their components across various capability levels.
\end{abstract}

	
	
\section{Introduction}
Digital twins, recognized as a pivotal technology in the realm of digitalization, harbor substantial potential for enhancing the operational efficacy of wind farms. This potential is manifested in heightened uptime, diminished capital and operational expenses, a reduced levelized cost of wind energy, and an augmented commitment to sustainability, as extensively deliberated. However, the precise definition of a digital twin has so far remained elusive. In a recent research endeavor~\cite{Stadtmann2023dti}, the authors have undertaken the task of elucidating the concept of a digital twin, presenting their common understanding, and embarking on the journey towards its realization within the ambit of the NorthWind project. This paper seeks to present some of the interesting results achieved in the project. The paper is organized into five distinct sections. Section~\ref{sec:Theory} expounds upon the shared comprehension of digital twins. Section~\ref{sec:Method} gives a brief introduction to the methodologies involved. The results and discussions are presented in Section~\ref{sec:Results}. Finally, the paper is concluded and future directions proposed in Section~\ref{sec:Conclusion}. 

\section{Digital twin} \label{sec:Theory}
\begin{figure}
    \centering             \includegraphics[width=\linewidth]{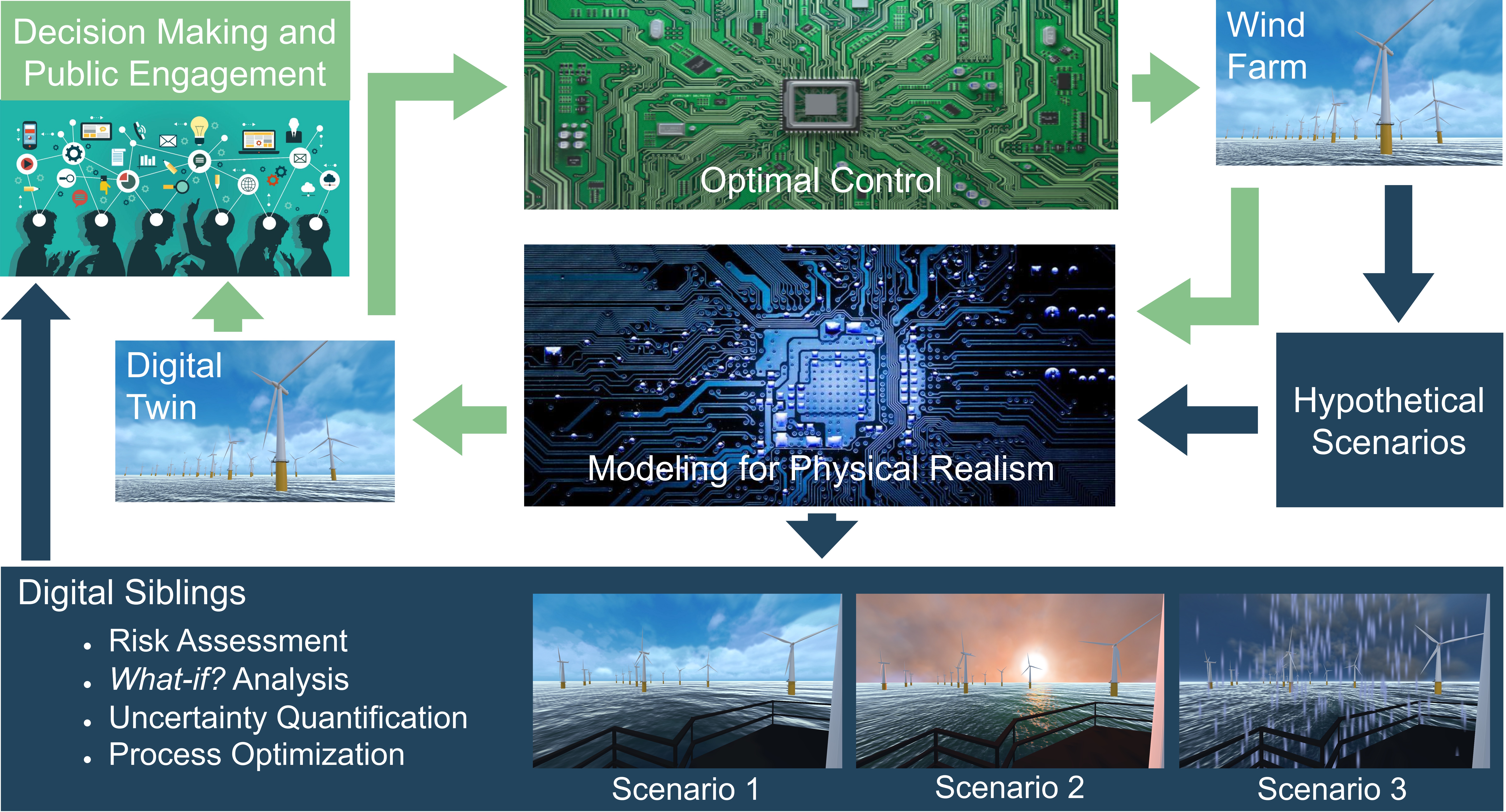}
    \caption{Digital twin concept}
    \label{fig:DT}
    \end{figure}

Following an extensive literature survey~\cite{Rasheed2020dtv} ``A digital twin is defined as a virtual representation of a physical asset or process enabled through data and simulators for real-time prediction, optimization, monitoring, control and improved decision-making.'' The concept is depicted in Figure~\ref{fig:DT}. The physical asset, in this case, a wind farm, is shown on the top right side. The state of the asset is measured in real-time through various sensors, resulting in a large data stream from the asset. Still, the data are coarse in spatial-temporal resolution and do not describe the physical behavior and future states of the asset. By exploiting physical models, the data can be complemented to instill physical realism into a virtual representation of the physical asset: the digital twin. This virtual representation can then be used to perform decision-making or even control the asset autonomously. Real-time data exchange in Figure~\ref{fig:DT} is marked with green arrows, while gray arrows and boxes describe the offline operations. Long-term decisions and recommendations can be provided by exploring hypothetical scenarios, e.g. for risk assessment or uncertainty quantification. This non-real-time application of the digital twin framework is also referred to as digital sibling. Note that while in the figure the virtual representation is labeled digital twin, data processing and analysis, physical models, decision support, and control are all implemented into the digital twin framework, and are therefore part of the digital twin.

    \begin{figure}
    \centering             \includegraphics[width=\linewidth]{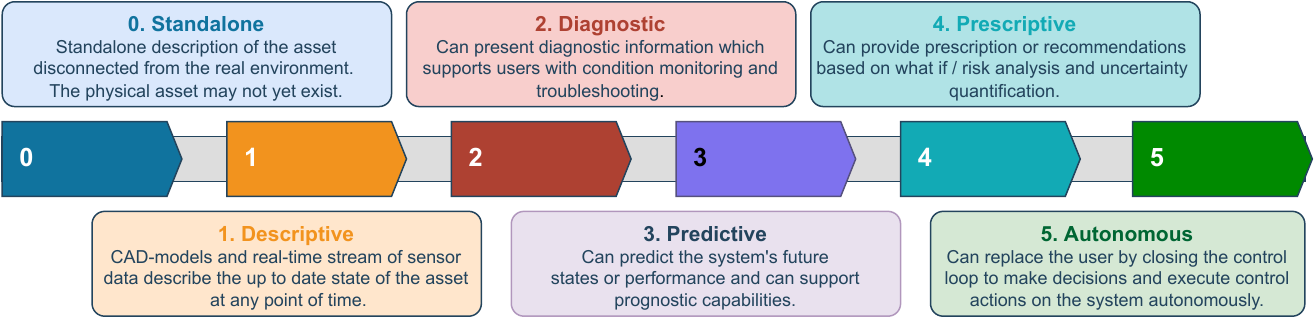}
    \caption{Capability levels of digital twins}
    \label{fig:CL}
    \end{figure}

To distinguish between the capabilities of digital twins (Figure~\ref{fig:CL}), they can be classified on a scale from 0-5 (0-standalone, 1-descriptive, 2-diagnostic, 3-predictive, 4-prescriptive, 5-autonomous)~\cite{Stadtmann2023dti} . The standalone digital twin is a virtual representation disconnected from the real asset and can be used during the design and planning stage. The descriptive digital twin describes the current state of the asset based on a real-time data stream. At the diagnostic level, the digital twin can be used to identify faults based on the current and historic asset state. In contrast to previous levels, the predictive digital twin also possesses the capability to forecast future asset states. The prescriptive digital twin provides recommendations to the user based on uncertainty quantification and risk analysis. Finally, the autonomous digital twin closes the loop as it optimizes the asset through bidirectional data exchange where the asset updates the digital twin and the digital twin controls the asset. More detailed explanations of the capability levels in the context of wind energy can be found in~\cite{Stadtmann2023dti}. The paper concludes that the primary challenges impeding the development and widespread adoption of the technology fall into three broad categories: data-, model-, and industrial acceptance-related challenges. The subsequent focus within the project has been on addressing these challenges, see the following section.
    
\section{Methodology} \label{sec:Method}
To illustrate the concept of a digital twin, the initial undertaking involved consolidating data, acknowledging that a realistic digital twin relies heavily on high-quality data. Data were curated from three distinct sources: (1) publicly available datasets (e.g., weather data), (2) data obtained from project collaborators, and (3) confidential data requiring anonymization. In addition, reverse engineering data (explained in Section~\ref{subsec:MethodreverseEngineering}) and synthetic data generated through numerical simulations (explained in Section~\ref{subsec:MethodSynthGen}) were also incorporated.

Subsequently, an asset information model was developed (Section~\ref{subsec:MethodRDS}), aimed at standardizing data to facilitate the scalability of digital twin technology. A data integration framework (Section~\ref{subsec:MethodDIF}) was then implemented to seamlessly fuse data from various sources, ensuring availability on demand while upholding data quality. This integrated data set was then utilized to construct computationally efficient and accurate models (Section~\ref{subsec:MethodComponentROM} to \ref{subsec:MethodGANS}).

The culmination of these efforts results in the creation of two reference wind farms, one onshore \cite{stadtmann_standalone_2023} and one offshore \cite{Stadtmann2023doa}. The methodology behind these endeavors is briefly outlined in the following subsections.

\subsection{Data acquisition and generation}
\subsubsection{Reverse engineering} \label{subsec:MethodreverseEngineering}

As previously noted, one of the challenges hindering the development of digital twins is stakeholders' reluctance to share data. Fortunately, there has been a positive shift, with many partners now beginning to share data. However, waiting for all the data to become available is not feasible. Thus, a reverse engineering approach \cite{Panjwani2024doa} was initiated based on the available resources.

For instance, wind turbine geometry can be reconstructed using turbine power curves and a baseline wind turbine available in the public domain. The methodology entailed a systematic process of analyzing power curves, formulating assumptions about control systems and airfoil geometries (chord and twist), and iteratively reverse engineering the turbine geometry. Control systems focus primarily on wind speed, rotational speed, and blade pitch to align with power and thrust performance curves. In the initial phase of this study, technical specifications were gathered from online resources, including the company homepage and third-party platforms, to establish a comprehensive dataset on wind turbine specifications. Critical parameters defining the study's scope, such as blade length, hub/rotor diameter, maximum and minimum rotor speeds, and cut-in/cut-out speeds, were identified. The SWT 2.3-93 wind turbine was chosen as the baseline, providing initial input and reference values. Subsequent systematic modifications to blade geometry and control parameters were introduced to explore design and operation variations.

NREL Aerodyn simulations were then performed to analyze the aerodynamic performance of the modified wind turbine configurations, aiming to predict power curves and other relevant data. Simulated results were compared with actual performance data from the initial wind turbine, facilitating a thorough assessment of the modifications' impact. An iterative refinement process followed, with adjustments based on comparative analysis, leading to optimized configurations.

The final phase involved validating the optimized wind turbine configurations through additional simulations and comprehensive comparisons with both the initial turbine and the actual SWT2.3-83 wind turbine.

\subsubsection{Synthetic data generation} \label{subsec:MethodSynthGen}
The purpose of synthetic data generation is to develop a better understanding of physics, as well as to use it to develop surrogate models. In the context of current work, two sets of simulated data were generated, one corresponding to solid mechanics and the other to fluid mechanics.  

\paragraph{Structural simulation}
The model is developed by DNV using Sesam GeniE.\footnote{Version 8.4-06} The wind turbine is not part of the model and in practice would be modeled using different software. Note that the K shape can be used as a building block to construct highly complex structures.

\paragraph{Flow around airfoils}
Simulation of 2D flow around two airfoils placed in tandem using OpenFoam was used to generate data by changing the Reynolds number. 

\begin{figure}
    \centering             
    \includegraphics[width=\linewidth]{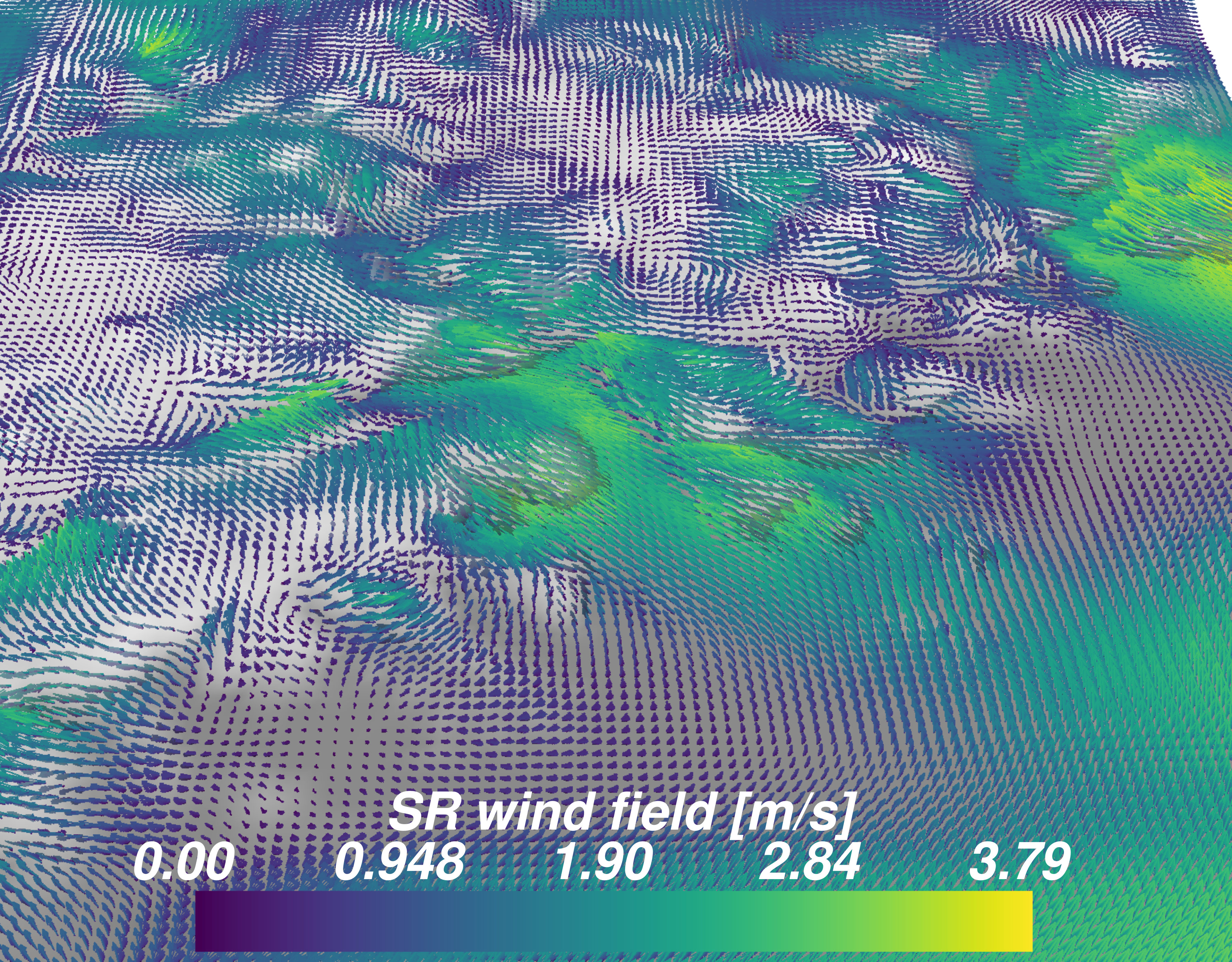}
    \caption{High resolution wind field in complex terrain}
    \label{fig:overview_SR}
    \end{figure}

\begin{figure}
    \centering             
    \includegraphics[width=\linewidth]{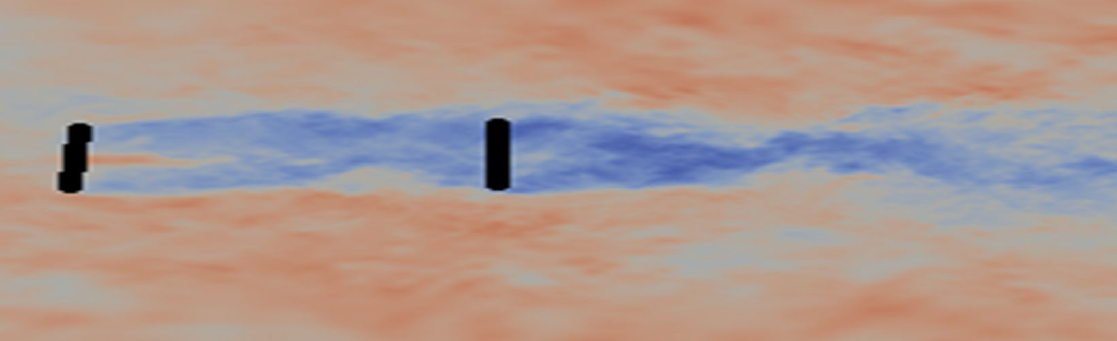}
    \caption{wakes-generated by turbine}
    \label{fig:wake-turbine}
    \end{figure}

\paragraph{High fidelity wind flow in complex terrain}   
The data for the Bessaker wind farm were generated using a HARMONIE-SIMRA multiscale wind modeling system (\cite{Rasheed2014amwa},\cite{Tran2020ges}). Wind forecast data from the HARMONIE model with a $2.5\,\mbox{km}\times2.5\,\mbox{km}$ horizontal resolution was used to force the SIMRA model which, in turn, generated a wind field at a fine horizontal resolution of $200\,\mbox{m}\times200\,\mbox{m}$ over a domain of $30\,\mbox{km}\times30\,\mbox{km}\times3\,\mbox{km}$. The simulation sample data can be seen in Figure~\ref{fig:overview_SR}. This data were used to train a Generative Adversarial Network model as explained in Section~\ref{subsec:MethodGANS}.
\paragraph{High fidelity wake simulation}
High-fidelity modeling of wakes behind wind turbines is crucial for providing accurate wake behaviour database and information for understanding and optimizing the wind turbine performance. The wakes impacts energy extraction efficiency in downstream turbines and influences structural integrity of turbines. For generating this database, the actuator line model (ALM) was coupled with Large Eddy Simulation (LES) turbulence model within the open-source Computational Fluid Dynamics (CFD) software OpenFOAM. In this study, ALM was employed to represent the turbine blades as discrete lines, allowing for a detailed representation of the rotor's aerodynamics. The LES method captures turbulent flow structures, providing a more accurate depiction of the complex wake dynamics. The use of OpenFOAM ensures flexibility and accessibility, enabling researchers to customize simulations according to specific turbine configurations. The methodology involved mesh generation, setting boundary conditions and initial conditions, and solver configuration for LES with ALM for an industrial scale reference NREL wind turbine. Here, the finest mesh was used in the region where the two turbines are located, with the finest mesh being about a tenth of the turbine diameter. Figure \ref{fig:wake-turbine} shows the resultant wakes generated in a two turbine set-up using the acutator line and LES methodology with the external wind flowing from left to right.

\subsection{Data standardization and management}  
    \begin{figure*}[h]
    \centering             \includegraphics[width=\linewidth]{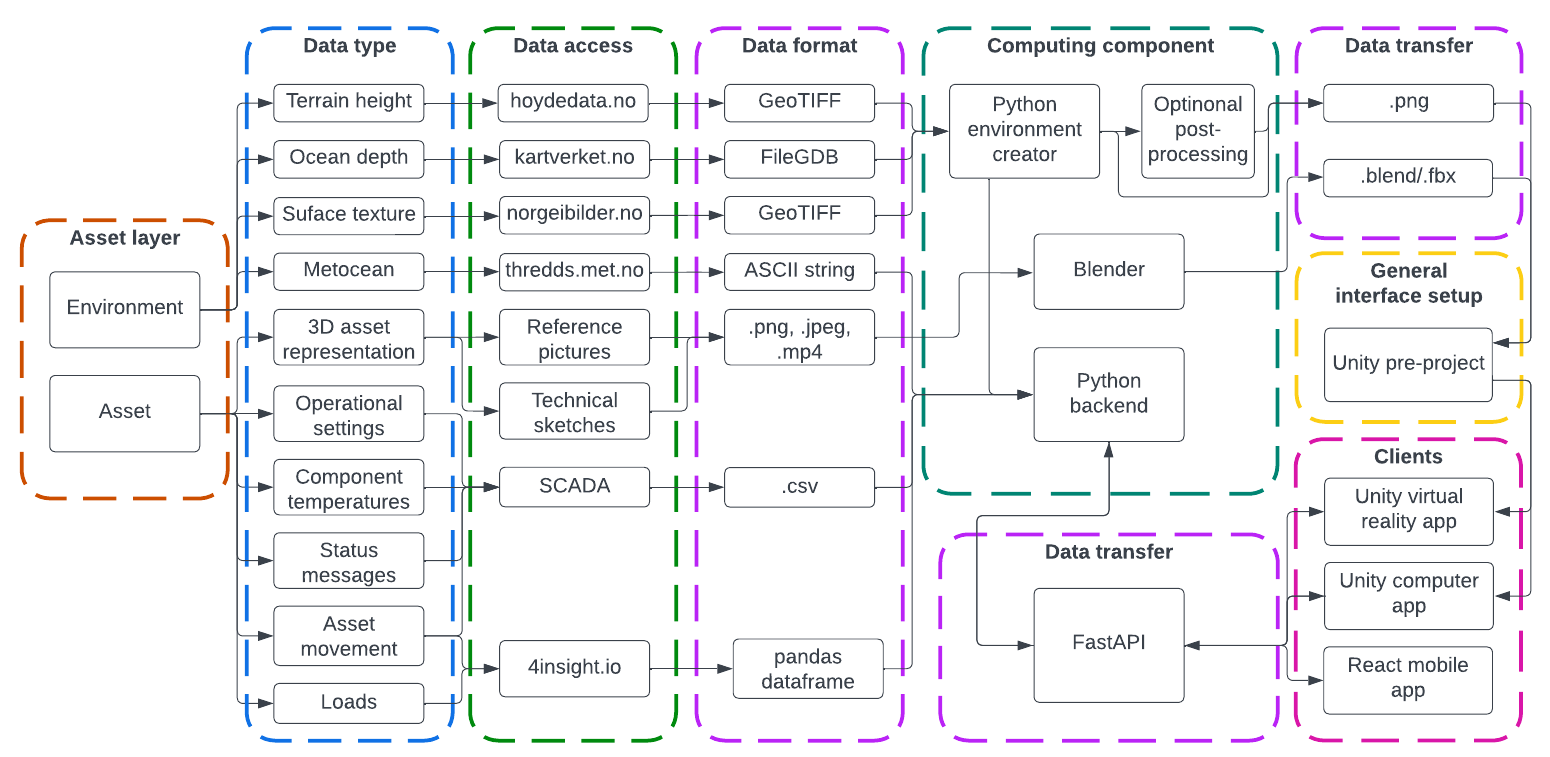}
    \caption{Data integration framework}
    \label{fig:DT_API}
    \end{figure*}
\subsubsection{Asset information model}\label{subsec:MethodRDS}

The simulation of wind turbines and the optimization of wind farms necessitate the integration of extensive meteorological, ecological, topological, structural, and operational data streams. To effectively structure and amalgamate diverse data types from various sources, each with distinct formats, syntax, and semantics (as outlined in Section \ref{subsec:MethodDIF}), a unified asset information model is indispensable. This model must be accessible and interpretable, offering a comprehensive depiction of the ``physical twin" (the asset) at a level of detail that suits the intended purpose of managing and correlating all data streams.

Drawing from the ISO 81346 standard for organizing systems (as cited in \cite{14:00_17:00i81} and \cite{14:00_17:00i82}), and its application in power systems \cite{14:00_17:00i81a}, ongoing efforts in standardizing wind turbines and foundations was extended to define a Reference Designation Structure (RDS) for various wind turbine generators and foundations. The RDS adopted a systems perspective of an overarching Wind Turbine Power System (PS), decomposed into a series of Technical Systems (TS) encompassing the structure (tower, nacelle, transition piece, and foundation), power production (blades, rotor, generator, converter, etc.), and various control, monitoring, and safety systems.

Further delineation within the Technical Systems led to Component Systems (CS), which encompass mechanical and electrical equipment, components within the primary and secondary structures, as well as monitors, sensors, and support systems. The ISO 81346 standard delineates how object systems can be defined from functional, product, and location aspects, encapsulating functional relations among object components, constructional relations among components, and spatial relations among components, respectively. These aspects of the RDS represent various asset structures and can be conceptualized as the framework(s) of a digital twin.

\subsubsection{Data integration framework} \label{subsec:MethodDIF}

Commencing from the descriptive level (capability level~1), digital twins are envisioned to synchronize seamlessly with their physical counterparts, providing an intricate representation of their current and historical states. This entails, firstly, aggregating data on the asset and its surroundings from various sources. Secondly, given the often sparse spatial and temporal dimensions of data, alongside the inability to measure certain parameters directly, models are indispensable for interpolating, extrapolating, and inferring data, including future values. Thirdly, digital twins necessitate the ability to present both data and analysis outcomes in an accessible manner across diverse clients and platforms.

A robust data integration framework is thus imperative, capable of accommodating a plethora of static and dynamic data sources from assets and environments, alongside modeling and analysis, and rendering visualizations across multiple client devices and platforms. Such a framework is exemplified in the context of offshore and onshore wind farms in \cite{Stadtman2023dif}, depicted in Figure \ref{fig:DT_API}, applicable across various digital twin domains. This framework encompasses static data such as terrain topology, surface characteristics, and asset geometry, as well as dynamic data like weather conditions, component temperatures, power generation, rotational speeds, strains, and status updates.

The framework leverages vendor Application Programming Interface (API) for data aggregation via a Python server, with potential utilization of cloud computing as an alternative to physical servers. Beyond data aggregation, the Python backend facilitates the integration of models written in diverse programming languages. Clients can access data through a cross-platform application developed using the Unity Engine, accessible across Windows, Linux, MacOS, Android, and iOS platforms. A RESTful application programming interface establishes the connection between the server and the application.

This data integration framework was demonstrated successful implementation in constructing digital twins for an onshore wind farm \cite{Stadtmann2023doa} and an offshore wind turbine \cite{stadtmann_standalone_2023}.

\subsection{Computaionally efficient modeling}
Ensuring real-time operation of a digital twin, models that are accurate and computationally efficient are required. Therefore, several models were developed and demonstrated with this criterion in focus.

\subsubsection{Component based reduced order models}\label{subsec:MethodComponentROM}
 Within the model order reduction approach, reduced-order modeling (ROM) has gained significant popularity~\cite{Fonn2019fdc,Quarteroni2014rom}. In ROM, full-order models (FOM) are projected onto a reduced-dimensional space, typically based on the proper orthogonal decomposition of the FOM simulation results (snapshots). If the information in the FOM results can be retained with a considerably reduced dimension, a ROM can achieve a speedup of several orders of magnitude relative to the FOM. However, a major issue with such ROMs arises from the degrees of parametric variation that may be present in complex systems like wind farms. With a very high-dimensional parameter space, the solution space becomes too varied, degrading the benefits of the reduced approach. To address this, the authors proposed considering a wind farm (or any other complex system) to be composed of components that are restricted in the parametric sense, making them amenable to ROM. For example, by assembling simple shapes such as airfoils and circles, a complete geometric model of a wind turbine / farm can be created. These shapes can then be modeled as opaque parametrized \emph{elements}, wherein only the external degrees of freedom are visible to the larger model, thereby reducing the total degrees of freedom from millions to hundreds. This idea is demonstrated through two examples, one corresponding to a solid mechanics simulation and another to a fluid mechanics simulation.  

\paragraph{Solid mechanics simulation}
The text describes a model of an offshore wind turbine (OWT) jacket foundation and tower structure. The structure consists of tubular steel beams and is attached to the seabed at $z=-27.5 \, \textrm{m}$. The jacket is $43.5\, \textrm{m}$ high, and the tower is $86.4 \, \textrm{m}$ high (the waterline is at $z=0$). Offshore jacket structures are often modeled using beam finite elements only. However, more refined modeling in certain regions, such as tubular joints, may be performed in order to improve numerical accuracy. Furthermore, for fatigue or ultimate limit-state post processing according to industry recommended practices, the conservatism of the computations may be reduced when shell models are considered. In the example here, the tubular joints between chords and braces are modeled using shell finite elements. Furthermore, eight of the tubular beam-element joints are replaced by three-dimensional shell-element reduced-basis components.

Each component has six ports, each with three translational and three rotational degrees of freedom. The degrees of freedom on the circular boundaries of the shell elements are condensed to a single point using explicit constraint equations. The finite-element discretization of each component consists of 26241 first-order triangular (three-noded) and quadrilateral (four-noded) linear shell finite elements. The resulting number of degrees of freedom is $153132$. Each component is parameterized by the wall thickness 
$\mu\in {\cal D} = [0.01 \, \textrm{m}, 0.1 \, \textrm{m}]$ of the top and bottom braces. The parameter enters into the equations as the thickness parameter in the shell finite element stiffness matrices. However, note that, due to the automatic affinization procedure, no detailed knowledge of the parameterization is needed. For the discretization of the remaining jacket structure, 102 third-order (two nodes with six degrees of freedom at each node) beam elements SESAM Sestra were utilized. A single beam element was employed between each joint, while the tower comprises multiple beam elements of varying diameters to approximate a conical shape. This section of the structure does not incorporate any parameters.

In the offline stage, Aroma was utilized to generate the component using 50 snapshots for each of the six ports and each of the six degrees of freedom on each port. The sampling parameters $\mu_i\in [0.01,0.1]$ are distributed according to the roots of the Legendre polynomials. Aroma invoked the Sesam Sestra as the full-order solver.

In the online stage, given a new parameter value $\mu$, Aroma is employed to generate the ROM stiffness matrix, which is automatically converted into a Sesam matrix superelement using a Python script. For simplicity, the same parameter value is used in all eight instances of the component. Leveraging standard Sesam superelement assembly functionality, each of the eight components is coupled with the remaining jacket structure. Consequently, an online solve of the size $288$, corresponding to the six degrees of freedom at each of the six ports of each of the eight component instances, is conducted.

\begin{figure}
    \centering             
    \includegraphics[width=\linewidth]{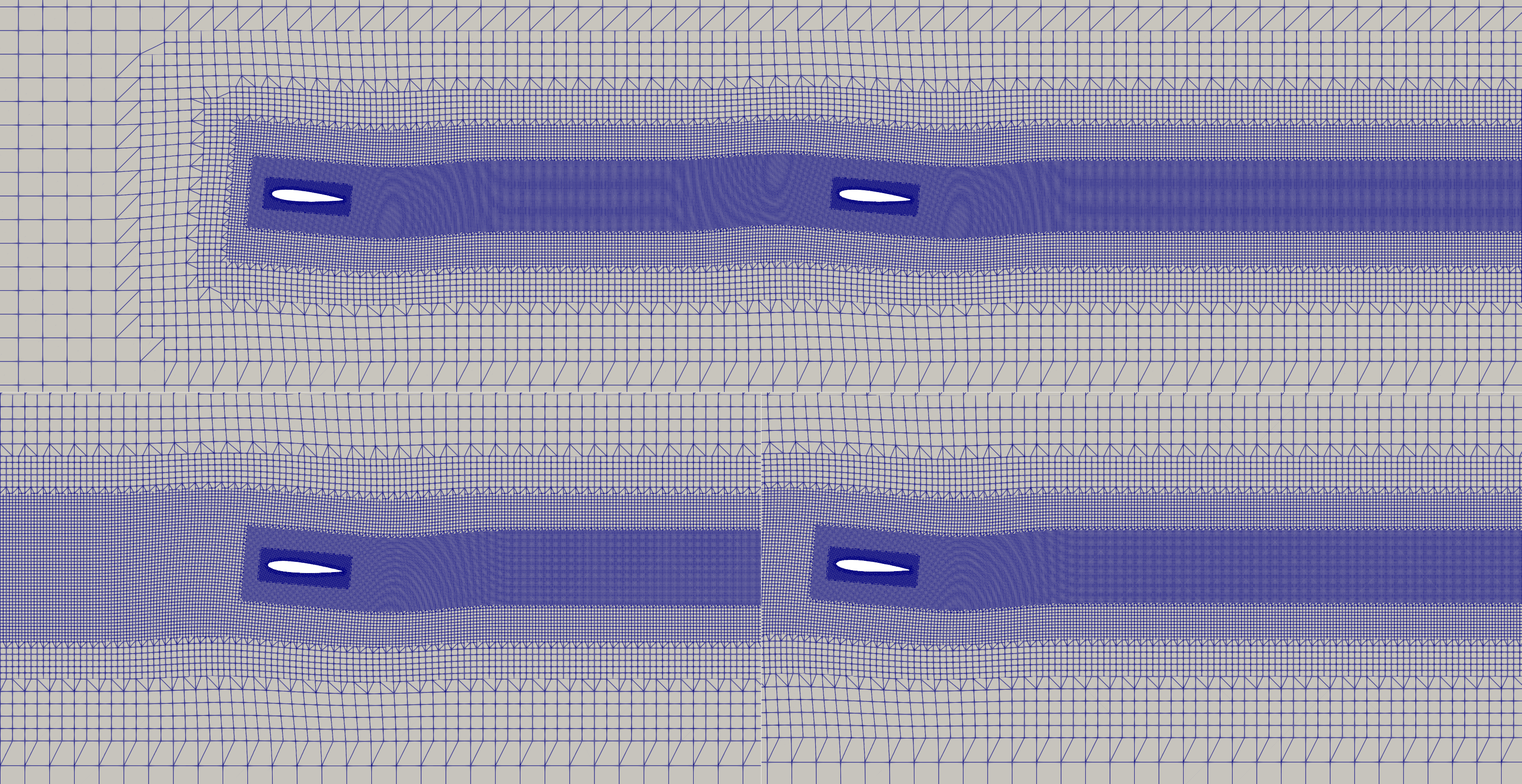}
    \caption{BENCHMARK AND ROM DOMAIN}
    \label{fig:Chain_ROM_mesh}
 \end{figure}

\paragraph{Fluid simulation}
In the proof-of-concept scenario, a pair of NACA 0012 airfoils is arranged in tandem, positioned $8$ chord-lengths apart. Both airfoils are set to an angle of attack of $\ang{5}$. The chord length is standardized to $1 \, \textrm{m}$, while the input velocity boundary condition is fixed at $0.1 \, \textrm{m/s}$, and the kinematic viscosity at $1\cdot 10^{-4} \, \textrm{m}^{2}\textrm{/s}$. Introducing a parameter, $\mu \in [0.1,1]$, scales the input velocity datum, thereby influencing the Reynolds number within the range $[100,1000]$. Two computational domains are created: the first, containing both airfoils, is discretized into $70,186$ cells, serving as a benchmark, while the second, comprising a single airfoil, consists of $45,931$ cells and constructs the Component-ROM component. Meshing is performed using OpenFOAM's snappyHexMesh utility, and a C++ code manages wing rotation and transition smoothing. Figure \ref{fig:Chain_ROM_mesh} visually illustrates this concept, presenting the full tandem domain at the top and two instances of the single-wing domain at the bottom, cut and overlapped to replicate the original configuration.
Utilizing OpenFOAM's SIMPLE algorithm, an ensemble of snapshots was gathered, starting with the upstream airfoil in free-stream. Solutions were obtained for three Reynolds numbers:$\, 100, 500, \, \textrm{and} \, 1000$. These snapshots served as the basis for formulating reduced bases for velocity and pressure, employing Galerkin projection. The free-stream ROM approximation was then utilized to establish accurate boundary conditions, replicating the downstream simulation. Subsequently, for the now-perturbed flow, an additional set of $10$ snapshots was collected, expanding the original sampling space. A new ROM was constructed, capable of approximating both upstream and downstream elements. 

\subsubsection{Windfarm layout optimization}\label{subsec:MethodWindFarmLayout}
In this work, the utilization of the open-source Julia code (FLOWFarm.jl, \cite{Thomas2022wec}) is demonstrated. The code is applied to the Unitech Zefyros virtual wind farm positioned at project area 3 for the Utsira Nord tender announced by the Norwegian Government in 2023. The requirement in the tender specified a minimum of $3.5 \, \mbox{MW/km}^{2}$ of installed capacity energy density for the farm in this area, which spans 
$179.3 \, \mbox{km}^{2}$. This leaves approximately $36 \, \mbox{km}^{2}$ of surplus area in relation to the government's energy density requirements. Assuming that the potential expansion into this surplus area is contingent on improvements in the network connection capacity, the entire area could be utilized at the specified density, resulting in an installed capacity of $627.5 \, \mbox{MW}$. Opting for the NREL $15 \, \mbox{MW}$ turbine as a reference, this translates to 41 turbines (or 125 turbines for $5 \, \mbox{MW}$ turbines). While acknowledging that not the entire area would realistically be utilized, the full area is employed for the purpose of comparing the two wake model codes. The process begins with the extraction of a wind rose for 100-m winds from ERA5 data, collected for every sixth hour from the year 2000 through December 2022. These data are sourced from a point located at $4.5 \, \mbox{E}$ and $59.1 \, \textrm{N}$ just offshore the west coast of Norway. To align with the chosen NREL turbine, the wind speed level is extended from $100 \, \mbox{m}$ to $150 \, \mbox{m}$ using a power law approach with a shear exponent of 0.15. The windrose bins are divided into $10^\circ$ (36 directional bins), with the average wind speed for each directional bin applied during optimization.

The windrose for various directions is depicted in Figure \ref{fig:layoutOptimization}. Analysis reveals that the flow is significantly influenced by the large-scale terrain of southern Norway, with the flow redirected from the southeast, deviating from the typical southwesterly direction due to the presence of mountains. In the ERA5-reanalysis, the land fraction at this point is less than 1\%, suggesting it may be regarded as an ocean point. To ascertain the layout with the maximum Annual Energy Production (AEP) for this site, 30 cases are initiated with turbines randomly positioned. Employing the Latin hyper-cubed technique ensures randomness while preventing turbine clustering or open spaces. Each of the 30 cases undergoes optimization with 70 iterations in three sequences, commencing from the randomly sampled turbine locations using a diffusive Bastankhah wake model (WES of 3, detailed in \cite{Thomas2022wec}). To prevent model crashes, a weak smoothing of the thrust and power curve is applied at weak winds.

\subsubsection{GANs}\label{subsec:MethodGANS}
\begin{figure}
    \centering             \includegraphics[width=\linewidth]{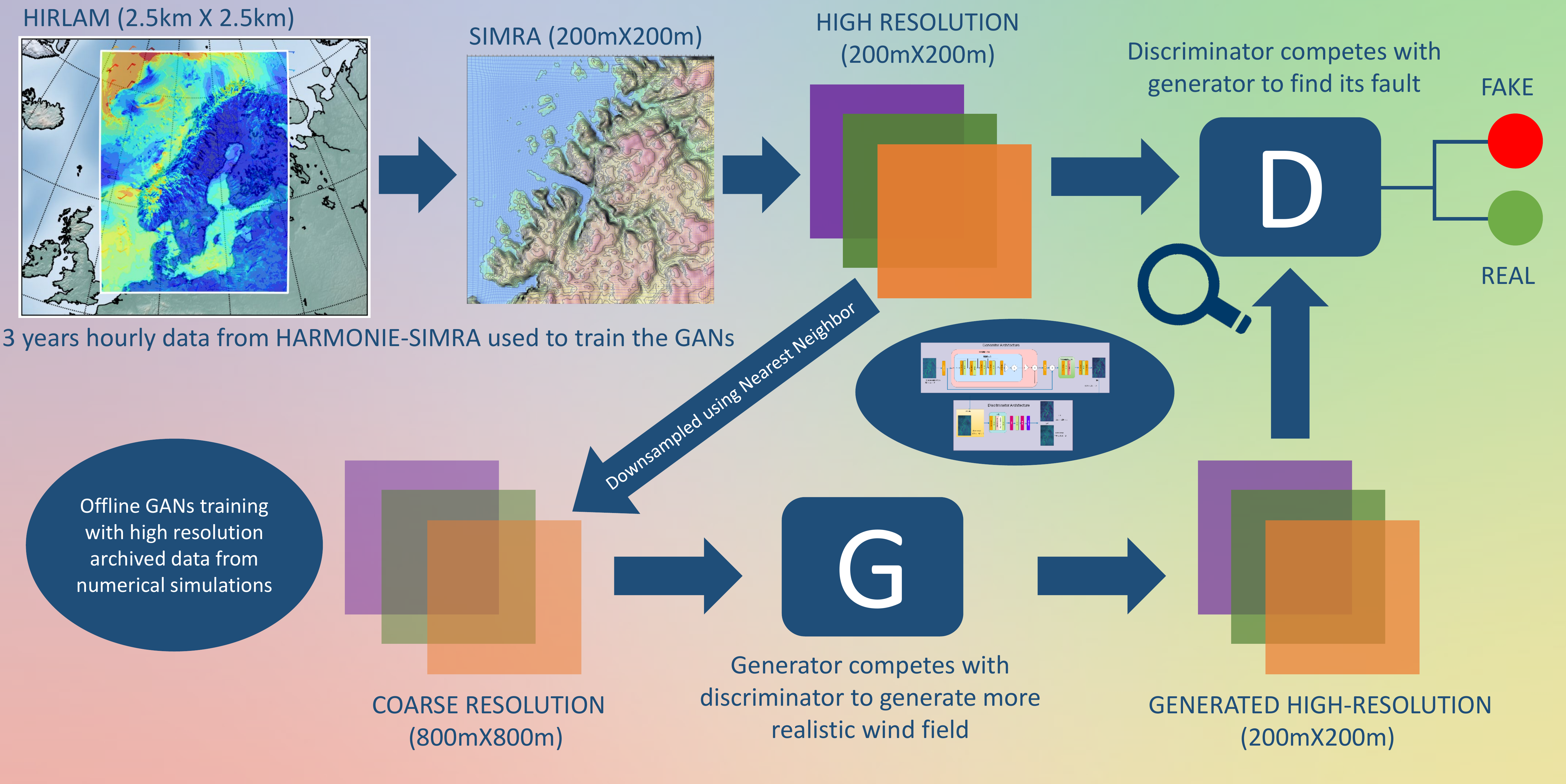}
    \caption{GANS framework}
    \label{fig:GANS_highlevel}
    \end{figure}
Atmospheric flows exhibit a diverse range of spatio-temporal scales, rendering real-time numerical modeling of turbulent flows in complex terrains at high resolution computationally impractical. The aim is to conduct simulations on a coarse mesh to enhance computational efficiency without compromising the accuracy achievable with a high-resolution mesh. To address this challenge, the utilization of generative adversarial networks (GANs) is proposed.

GANs consist of two antagonistic neural networks—the generator (G) and the discriminator (D) (see Figure~\ref{fig:GANS_highlevel}). The generator, given an input, produces an output, while the discriminator assesses whether the output appears genuine or artificial. Through joint training, the generator becomes proficient in generating results that closely resemble real-world scenarios. In the current study, GANs are trained by inputting coarse-scale simulations corresponding to a horizontal resolution of $800 \, \mbox{m} \times800 \, \mbox{m}$ and compelling the network to generate a high-resolution field with a resolution of $200 \, \mbox{m} \times 200 \, \mbox{m}$. The $800 \, \mbox{m} \, \times800 \, \mbox{m}$ resolution is derived by applying a nearest neighbor algorithm to the synthesized data with a $200 \, \mbox{m} \times 200 \, \mbox{m}$ resolution. What distinguishes this work from prior efforts in image resolution upscaling is the incorporation of a cost function that adheres to the principles of physics. More details of the methodology can be found in~\cite{Wold2023mlf}.

\subsection{Virtual and mixed reality for informed public engagement and decision making}
Virtual and mixed reality technologies have significantly advanced the field of digital twins, offering immersive and interactive experiences that enhance the understanding and management of physical entities. In the context of digital twins, virtual reality (VR) enables users to visualize and explore virtual replicas of real-world objects or systems in a simulated environment. This allows for a more intuitive comprehension of complex structures, such as buildings, machinery, or entire urban landscapes. Mixed reality (MR), on the other hand, combines virtual elements with the real-world environment, creating a seamless blend of digital and physical information. This integration is particularly valuable in scenarios where real-time data overlays onto physical counterparts, offering users a holistic perspective and facilitating decision-making processes. In the NorthWind project, extensive use of this technology is made for better communication of the results achieved.
    
\section{Results and Discussion} \label{sec:Results}
\subsection{Reverse engineered data}
    \begin{figure}
    \centering             \includegraphics[width=\linewidth]{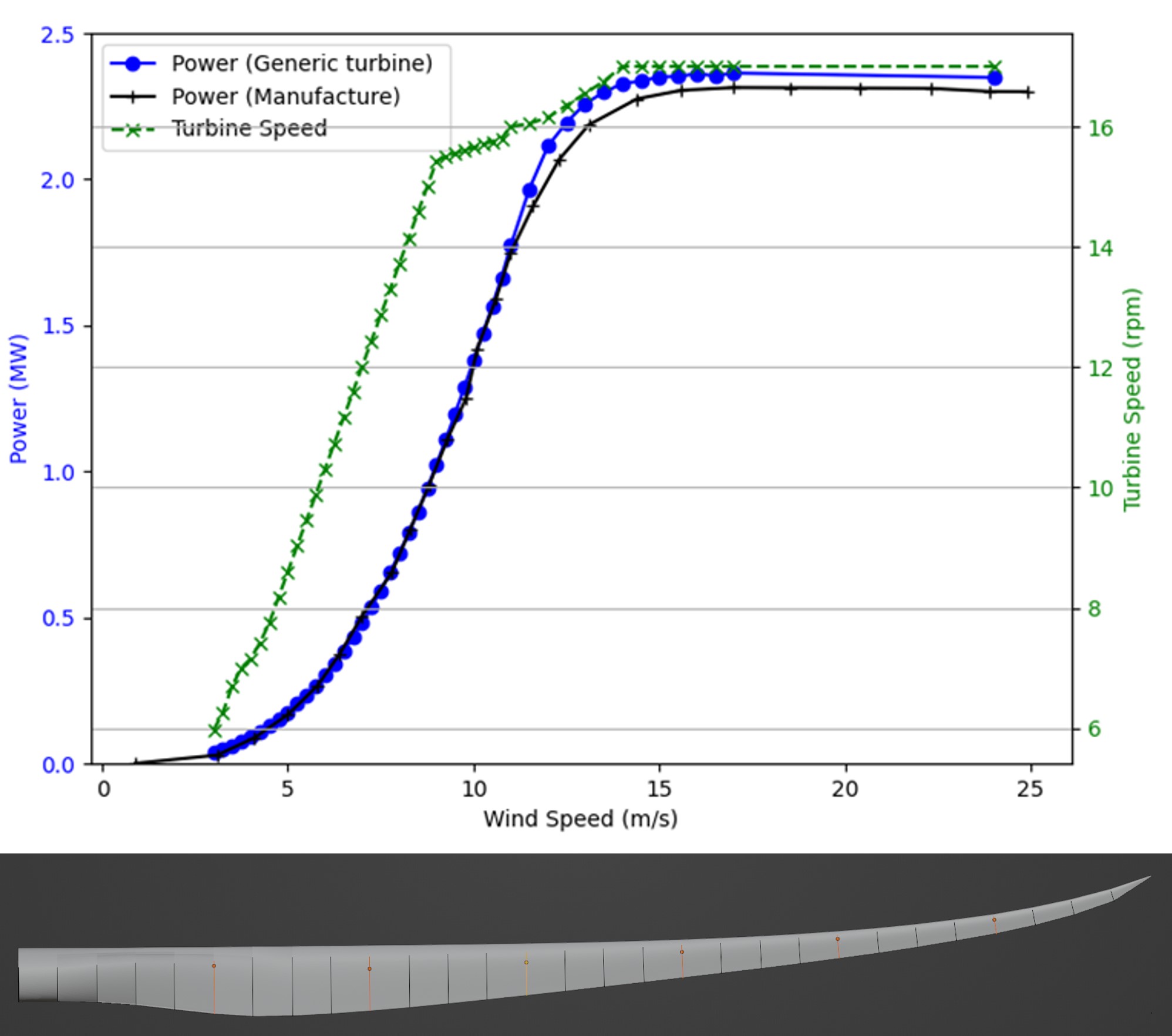}
    \caption{Reverse Engineered Blade}
    \label{fig:reverseEngineering}
    \end{figure}
Designing a wind turbine control system is a multifaceted task considering numerous variables to ensure the turbine is efficient, safe, and reliable. A simplified control system has been developed primarily centered around wind speed, rotor speed, and blade pitch. The development of this simplified control system involved a series of Aerodyn simulations, with pre-described airfoil properties. The starting point for this control system was the SWT 2.3-93 wind turbine. Throughout the Aerodyn simulations, adjustments were made to both the turbine speed and blade pitch to achieve a power curve of the SWT 2.3-83 wind turbine. Utilizing the developed control system and blade aerodynamic geometry, Aerodyn simulations of the turbine were conducted. The simulated power curve compared with the actual power curve is presented and shown in Figure~\ref{fig:reverseEngineering}. The reconstructed turbine geometry showed a good comparison with the anticipated power output across varying wind speeds. Through an iterative process, turbine geometry was successfully built that closely reflected the expected performance characteristics. The studies showed that the reverse engineering approach can effectively be used to design wind turbine geometry, which signifies a significant milestone in developing a digital twin for the turbine. However, while these results are promising, further validation and refinement are essential for enhancing the accuracy and applicability of this methodology, particularly under diverse operational conditions and environmental factors. 
    \begin{figure}
    \centering             
    \includegraphics[width=\linewidth]{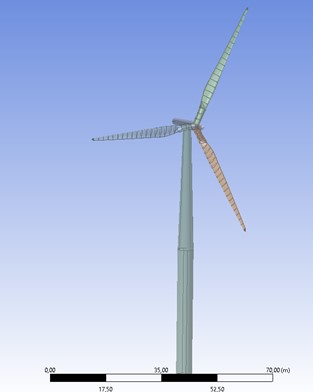}
    \caption{Reverse Engineered Wind Turbine}
    \label{fig:reverseEngineeingCAD}
    \end{figure}

The CAD model of the wind turbine was constructed using CAD software and is shown in Figure~\ref{fig:reverseEngineeingCAD}. The blade geometry, hub radius, tower dimensions, and airfoil profiles defined above were integrated to form a comprehensive and accurate digital model. The CAD model serves as a visual and analytical tool for assessing the wind turbine's design and performance.

\subsection{Asset information model} \label{subsec:AIM}
\begin{figure}[h]
    \centering             
    \includegraphics[width=\linewidth]{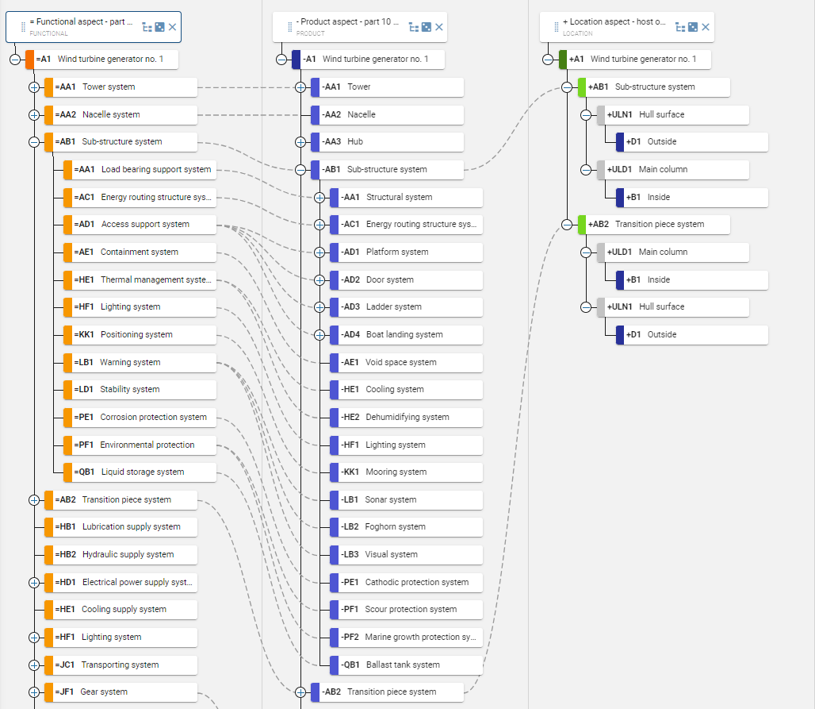}
    \caption{The Functional, Product and Location aspect of the Wind turbine and foundation RDS}
    \label{fig:AIM}
    \end{figure}

Figure~\ref{fig:AIM} depicts part of a combined wind turbine and foundation RDS with Functional, Product, and Location aspects, along with some of the relations between them. In Northwind, the RDS shown in Figure~\ref{fig:AIM} was utilized to develop a set of templates for a selection of fixed and floating asset types (jacket structures, monopiles, gravity-based structures, semisubmersibles, and spar buoys). In ongoing work, a model (RDS) of the Zephyros Spar at the Norwegian Metcenter is employed to connect data streams from sensors to the Functional aspect RDS (indicating the type of data delivered by the various sensors) and represent the sensors in the Product aspect (specifying the type of sensor they are). The Location aspect is then utilized to connect the sensors to their position in or on the structure to obtain a "complete" model of available data, indicating which sensors deliver it and where on the structure it originates.

\subsection{Component ROM}

\begin{figure}
    \centering             \includegraphics[width=\linewidth]{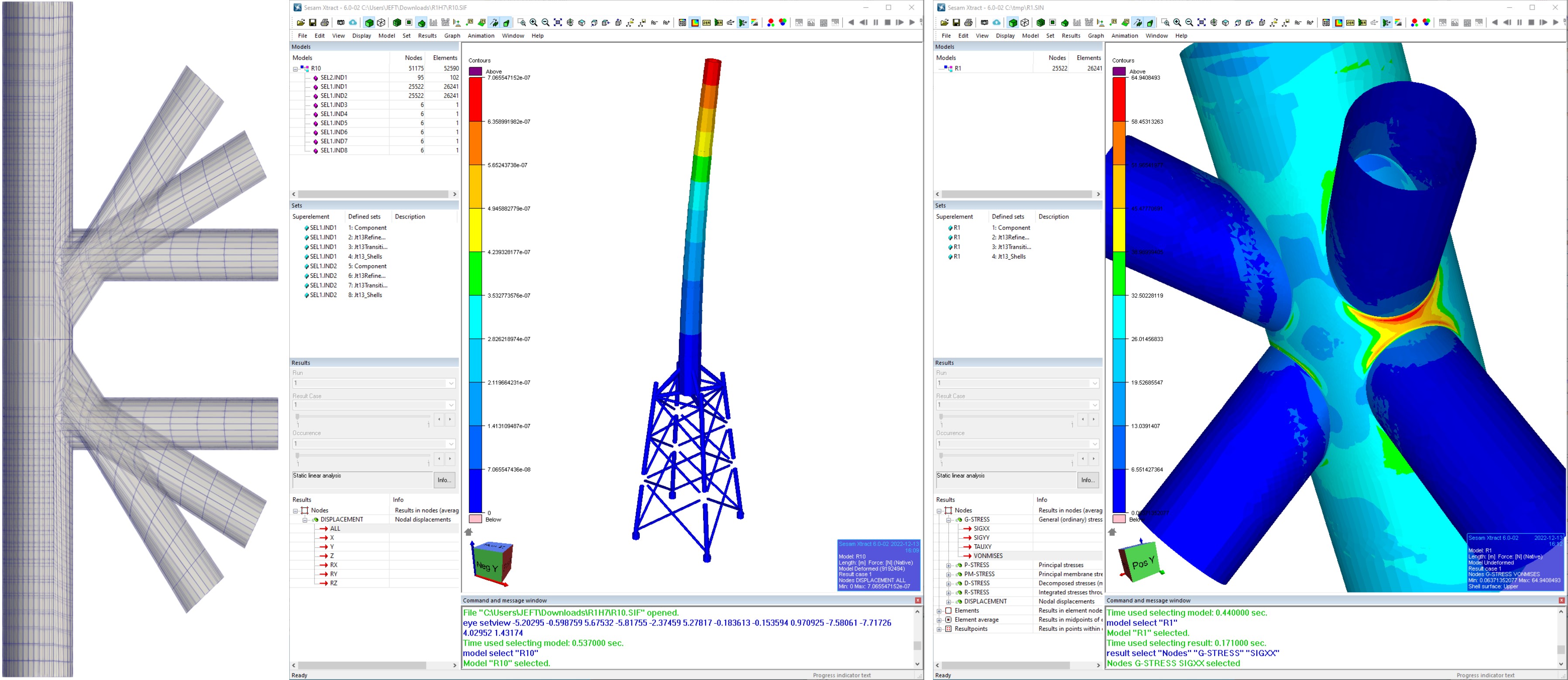}
    \caption{Component ROM solid mechanics simulations}
    \label{fig:componentROM}
    \end{figure}

\paragraph{Solid mechnics}
Performing the online stage, i.e., assembly and solution of the statically condensed system of equations of size 288, using the reduced-order model, takes 6.9 seconds. If, additionally, reconstruction of the displacement field in the shell elements is performed, the time increases to 12.4 seconds. In comparison, assembly and solution of the statically condensed system of equations of size 288, using the high-fidelity model, takes 72.1 seconds on our laptop computer\footnote{Using Sestra 8.8-02 \label{us882}} Note that this requires the factorization of a matrix of size $N_h=153132$ in order to perform the static condensation. Solving the global finite-element system without static condensation results in a system of equations of size $N_h=1222380$ and a solution time of $471$ seconds.\textsuperscript{\ref{us882}} Although the reduced basis and high-fidelity model is solved using different software (aroma and Sestra, respectively), the timings do give an indication that significant speedup is achievable. Reslts from the simulations are shown in Figure \ref{fig:componentROM}.

\begin{figure}[h]
    \centering             
    \includegraphics[width=\linewidth]{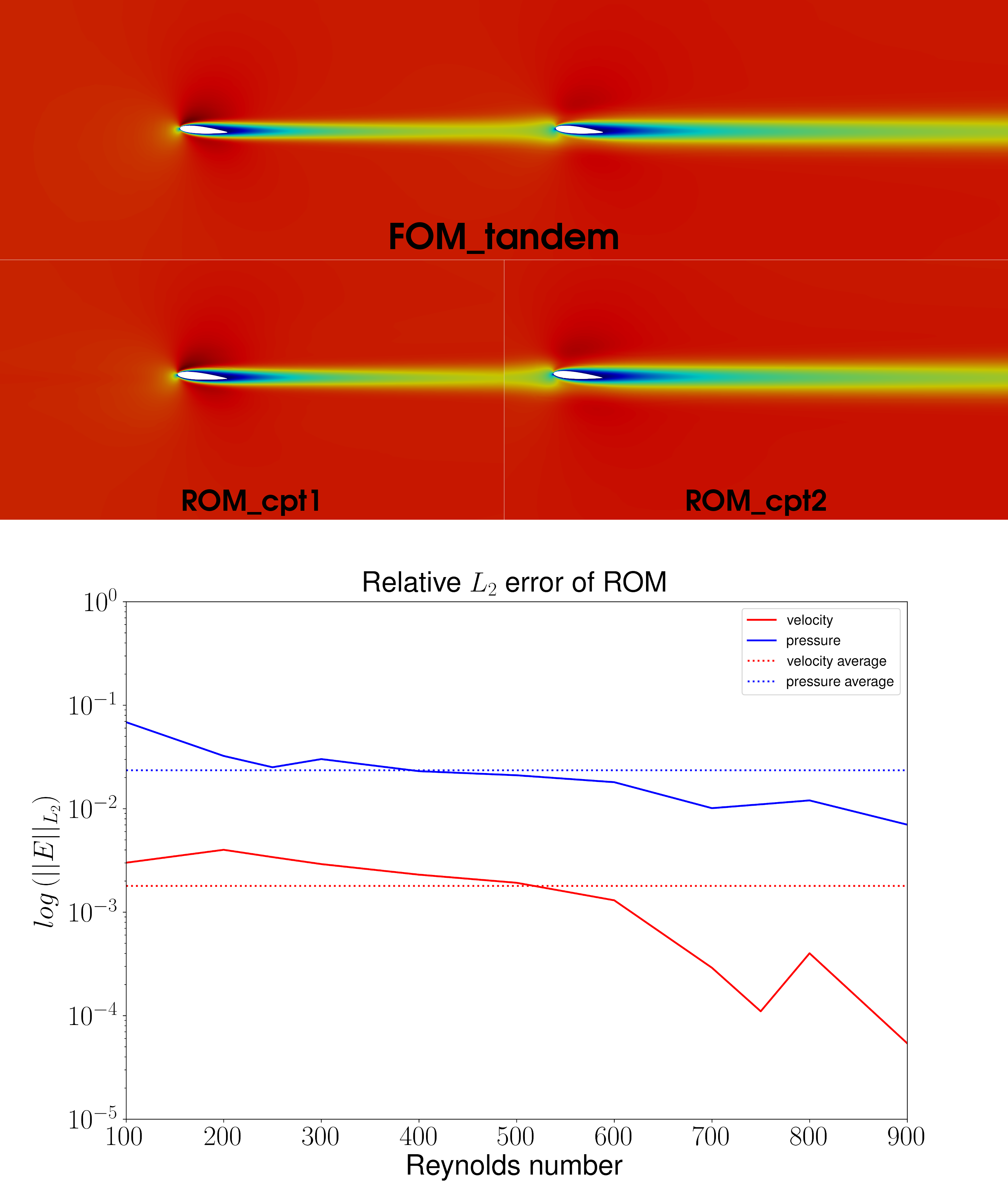}
    \caption{Component ROM APPROXIMATION FOR REYNOLDS NUMBER 750 AND PROJECTION ERROR VS REYNOLDS NUMBER}
    \label{fig:Chain_ROM_approx}
\end{figure}

\paragraph{Fluid mechanics}
Figure \ref{fig:Chain_ROM_approx} illustrates that with the constructed ROM, the flow around both elements can be approximated with great accuracy, achieving approximately $0.2\%$ and $2\%$ relative $L^2$ error for velocity and pressure, respectively. Similarly, the force coefficients can be efficiently approximated. The achieved speed-up, using just $6$ degrees of freedom for velocity and $4$ for pressure, is approximately $15,000$, indicating promise for larger systems.
 
\subsection{Windfarm layout}
\begin{figure}[h!]
    \centering             \includegraphics[width=\linewidth]{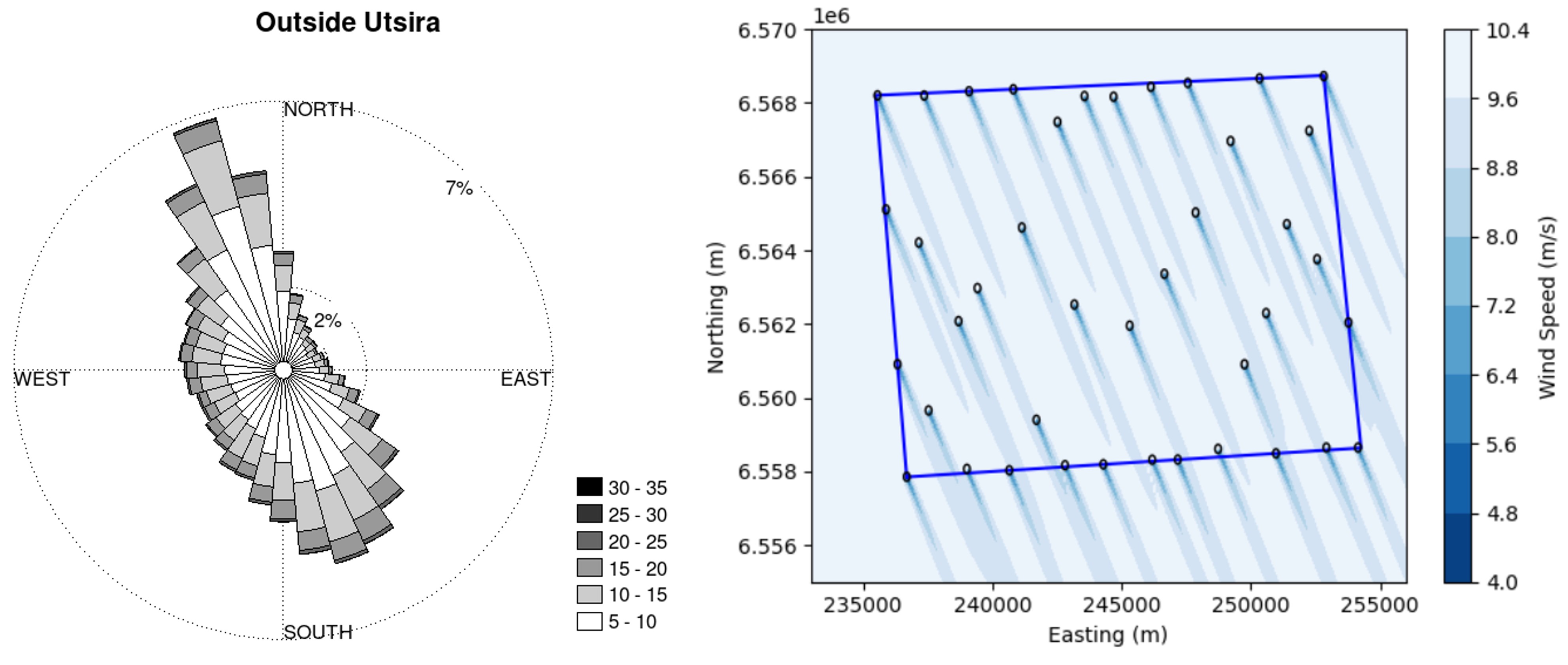}
    \caption{ Left: Windrose of the virtual farm's location. Right: An example of the flow field for the north-northwesterly predominant wind direction for the optimized layout.}
    \label{fig:layoutOptimization}
    \end{figure}
The proposed layout for 41 15MW turbines is shown in Figure~\ref{fig:layoutOptimization}. We see that the distribution has some clustering along the northern and southern edges which would be the sides normal to the main wind direction. In assessing the FLOWFarm optimization tool, a comparison with the commercial tool WindPRO1 was conducted to enhance confidence. The problem was configured similarly, considering area restrictions, inflow conditions, and wind turbine specifications, albeit using the Jensen wake model in WindPRO instead of the Bastankhah model. The wind farm layout was optimized in WindPRO, revealing a preference for turbine placements along the northern and southern borders, akin to the FLOWFarm results. Power production comparison between the two layouts, facilitated by importing FLOWFarm solutions into WindPRO, demonstrated virtually identical results. Notably, the optimization time differed significantly, with WindPRO taking hours and FLOWFarm mere minutes, while maintaining low wake losses (3.5\%).

\subsection{GANs enabled microscale windflow simulations}
    \begin{figure}[h]
    \centering             \includegraphics[width=\linewidth]{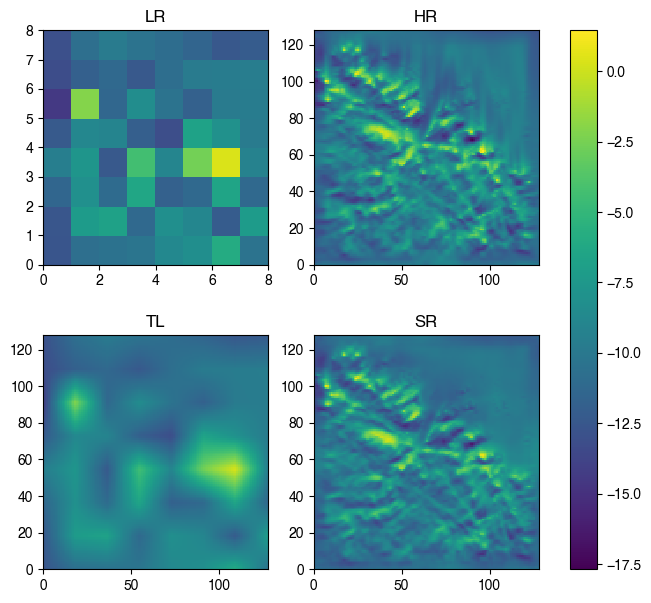}
    \caption{GANS-based superresolution simulation}
    \label{fig:results_gans}
    \end{figure}
Figure~\ref{fig:results_gans} displays the outcomes of the GANs-based reconstruction of the high-resolution wind field. The figure comprises four subfigures (LR, HR, RL, SR). LR represents a coarse-scale simulation $(800 \, \mbox{m} \times 800 \, \mbox{m})$ of the wind field in complex terrain, while HR corresponds to a horizontal resolution of $200 \, \mbox{m} \times 200 \, \mbox{m}$. TL denotes the trilinear interpolation of the LR field on a mesh corresponding to HR, serving as the ground truth. SR illustrates the upscaled resolution of the wind field achieved through the GANs network. It is clear that the SR field bears a striking resemblance to the one obtained through trilinear interpolation. The integration of GANs facilitates a computational speedup of up to 100 times without compromising accuracy. Additional details about this work are available in our archived article \cite{Wold2023mlf}.
\subsection{Informed public opinion and decision making}
\begin{figure}
    \centering             \includegraphics[width=\linewidth]{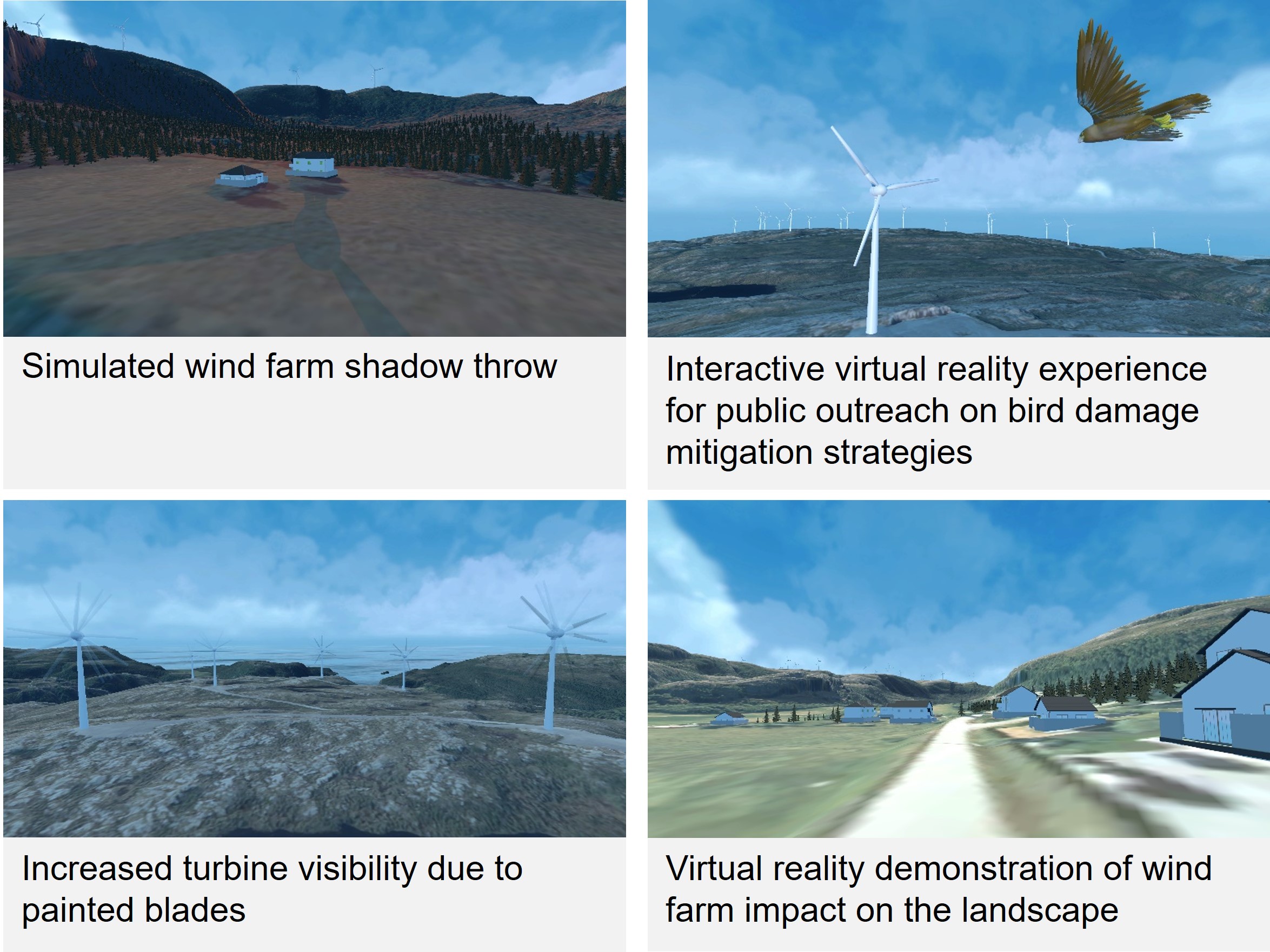}
    \caption{Informed Public Engagement and decision making}
    \label{fig:InformedPE}
    \end{figure}

Digital twins with advanced visualization and communication features are indispensable tools in the realm of wind energy, offering significant advantages in understanding and managing the impact of wind farms on both human communities and the surrounding ecosystems. In terms of shadow analysis, a digital twin can simulate and optimize the shadow patterns cast by wind turbines, allowing the strategic placement and design of these structures to minimize adverse effects on nearby residents. In addition, when it comes to wildlife conservation, digital twins can model and predict the ecological consequences of wind farm development, providing insight into how shadows can influence natural habitats. The innovative practice of painting wind turbine blades black and integrating sound generators, as simulated by the digital twin, becomes particularly relevant for mitigating visual and auditory disturbances, thereby fostering a more wildlife-friendly energy infrastructure. Additionally, the capacity of the digital twin to simulate changes in visibility due to wind farm development ensures that local communities are well informed about alterations in their surroundings, promoting transparency and community engagement in sustainable wind energy initiatives. Figure~\ref{fig:InformedPE} gives some illustrations of the digital twin framework that has been developed for one of the existing onshore wind farms.

\section{Conclusion and Future Work} \label{sec:Conclusion}
In the current work, the authors present some of the highlights of the research conducted within the NorthWind project. The presented work can be summarized as follows

\begin{itemize}
    \item An asset information model was developed to enhance the scalability of the technology.
    \item Data from various sources were collected and consolidated to create digital twins, with particular focus on employing reverse engineering for necessary data and establishing a well-structured data flow pipeline.
    \item Models for windfarm layout optimization were introduced, utilizing a gradient-based approach to achieve simulation times in the order of minutes, a significant enhancement compared to state-of-the-art approaches with simulation times in the order of hours or days.
    \item Generative Adversarial Networks (GANs) were presented for upscaling the resolution of wind fields, attaining the accuracy of a $200 \, \mbox{m} \times 200 \, \mbox{m}$ simulation at the computational cost of $ 800 \, \mbox{m} \times 800 \, \mbox{m}$ resolution simulations, resulting in a 100x speed-up of high-resolution wind simulations in complex terrain.
    \item A component-based Reduced Order Model (ROM) was introduced, demonstrating how online computational efficiency can be achieved at the expense of expensive offline simulations. It was also illustrated how complex assets can be constructed using ROMs of much simpler building blocks, achieving a speed-up on the order of $10^{4}$.
    \item An elegant and intuitive method for communicating results to a diverse range of stakeholders was presented for informed decision-making and public participation.
\end{itemize}

Future work envisions enhanced utilization of the asset information model in numerical simulations and control processes, optimizing its effectiveness. To achieve this, the adoption of evolutionary algorithms for reverse engineering is proposed, aiming for a more efficient alternative to the current brute-force approach. The windfarm layout optimization process is anticipated to evolve by incorporating additional factors and constraints, further refining the models for enhanced decision-making. Additionally, the development of Component-ROMS for fluids and fluid-structure interactions is identified as a key focus for future research. It is important to highlight that although the authors presented high fidelity wake simulation data, it has not yet been used for building Component-ROMS. Once such a component-ROM is operational, it can be utilized within the wind farm layout optimization work presented in this study. 

In the future, efforts will be directed towards expanding the capabilities of GANS to model the marine boundary layer, with a particular focus on understanding the impacts of stratification. Additionally, there is a plan to seamlessly integrate the data flow framework, asset information model, simulations, and control algorithms. This comprehensive integration is expected to yield an intuitive human-machine interface, offering a robust platform for efficient decision-making. This integration process will improve the overall functionality and user experience, thus contributing to advancements in wind energy technology.

\section*{Acknowledgments}
This publication has been prepared as part of NorthWind (Norwegian Research Centre on Wind Energy) co-financed by the Research Council of Norway (project code 321954), industry, and research partners. More details are given at \url{www.northwindresearch.no}.
Data for the project realization was provided by SINTEF Energy Research.

\end{document}